\documentclass[aps,prl,twocolumn,groupedaddress,showpacs]{revtex4}

\usepackage{graphics}
\begin{document}




\title{Andreev reflections on $Y_{1-x}Ca_{x}Ba_{2}Cu_{3}O_{7 - \delta }$ evidence for an unusual proximity effect}



\author{A. Kohen}
\email[]{kohen@tau.ac.il} \homepage[]{www.tau.ac.il\~supercon}
\author{G. Leibovitch}
\author{G. Deutscher}

\affiliation{School of physics and Astronomy, Raymond and Beverly
Sackler Faculty of Exact Science, Tel-Aviv University, 69978
Tel-Aviv, Israel}

\date{\today}

\begin{abstract}
We have measured Andreev reflections between an Au tip and
$Y_{1-x}Ca_{x}Ba_{2}Cu_{3}O_{7 - \delta }$ thin films in the
in-plane orientation. The conductance spectra are best fitted with
a pair potential having the $d_{x^{2}-y^{2}}$\textit{+is
}symmetry. We find that the amplitude of the \textit{is }component
is enhanced as the contact transparency is increased. This is an
indication for an unusual proximity effect that modifies the pair
potential in the superconductor near the surface with the normal
metal.

\end{abstract}

 \maketitle

 The superconducting state is characterized by a pair potential (PP) $\Delta \left( r \right) =
V\left( r \right)F(r)$ where V(r) is the potential representing
the electron-electron interactions and $F\left( r \right) \equiv
\left\langle {\Psi _ \uparrow \left( r \right)\Psi _ \downarrow
\left( r \right)} \right\rangle $ is the probability amplitude for
finding a Cooper pair. In the case of a contact between a normal
metal (N) and a superconducting metal (S), far from the interface
we expect F(r) to be zero in N and to reach a constant value in S.
However near the interface F(r) has a non zero value in the normal
metal due to electron pairs leaking from S into N. This phenomenon
is known as the proximity effect and has been studied both
theoretically and experimentally for the case of superconductors
with a pair potential having an s-wave
symmetry\cite{Deutscher:1969}. Microscopically the mechanism
involved in the creation of this non zero pair amplitude F(r) in
the normal metal, is a special reflection process known as Andreev
reflection\cite{Andreev:1964}, which occurs when there is an
abrupt change in the PP. Such a change occurs at a clean N/S
contact, where F(r) is continuous, due to the difference in V(r)
in the two materials. An electron approaching the
superconductor(SC) from the normal metal with energy smaller than
$\Delta $ cannot enter as a quasiparticle into the superconducting
condensate. Instead the electron is reflected as a hole and a
Cooper pair is added to the condensate. Therefore by measuring
Andreev reflections one can investigate the properties of the
superconducting PP.

For a perfectly transparent, small N/S contact, of size $a<<l$,
where $l$ is the mean free path, the Andreev reflection process is
manifested by a low bias ($eV\leq\Delta$) conductance which is
twice as large in comparison to the high bias one. Blonder
\emph{et al.}\cite{Blonder:1982}(BTK) have calculated the
conductance of an N/S contact with an additional barrier
represented by a delta function potential $U\left( x \right) =
H\delta \left( x \right)$. They have defined a dimensionless
parameter representing the barrier strength, $Z = \frac{H}{\hbar
v_F }$, where $v_{F}$ is the Fermi velocity. A clean N/S contact
is described by Z=0 and a high barrier tunnelling contact by
$Z>>1$. For finite Z values, the conductance is depressed at low
bias and enhanced at $eV=\Delta$. In the case of high temperature
superconductors (HTS) various experiments have indicated that the
PP has a $d_{x^{2}-y^{2}}$ (\emph{d}) symmetry\cite{Tsuei:2000},
described by $\Delta \left( k \right) = \Delta _0 \cos \left(
{2\theta } \right)$, where $\theta$ is the polar angle measured
from the crystallographic a-axis. This PP is very different from
the isotropic s-wave PP as it changes sign and has nodes for
$\theta = \frac{\pi }{4} + \frac{\pi }{2}n$, where n is an
integer. Thus it is expected that the properties of N/HTS contacts
should differ from those of N/S contacts. Tanaka and
Kashiwaya\cite{Kashiwaya:1995} have extended the BTK calculation
to the case of an anisotropic PP, $\Delta(\theta)$ and have shown
that the conductance curves differ from those calculated by BTK.
They have defined $Z=\frac{2mH}{\hbar^{2}k_{F}}=2Z_{BTK}$
henceforth we shall use this definition. For a \emph{d} PP a zero
bias conductance peak (ZBCP) appears for in plane contacts
reflecting the existence of Andreev surface bound
states\cite{Hu:1994}. It is most pronounced  for (110) contacts.
For a (100) contact, the low bias dip which evolves in the case of
a s-wave PP for $Z>0$, appears only for $Z\gtrsim0.4$ and the
amplitude of the normalized conductance maxima is lowered
considerably, reaching a value of around 1.5, as shown in Fig. 1a,
compared to a value of around 2.1 in the s-wave case for this Z
value. Experimentally, data on Andreev reflection at low Z
(Z$<$1)contacts with HTS is limited. Results on YBCO were reported
by Hess \emph{et al.}\cite{Hass:1992} and Yagil \emph{et
al.}\cite{Yagil:1995}. Wei \emph{et al.}\cite{Wei:1998} have
reported a measurement on YBCO with a fit to a \emph{d} PP.
Measurements on La$_{2 - x}$Sr$_{x}$CuO$_{4}$ were reported by
Achsaf \emph{et al.}\cite{Achsaf:1997} with a fit to an
anisotropic \textit{s-wave} PP and by Gonnelli \emph{et
al.}\cite{Gonnelli:1} with a fit to a
$\emph{d}$\textit{+id}$_{xy}$ PP.

Here we report on Andreev reflection spectroscopy measurements
using a point contact between a normal metal (Au) and $Y_{1 -
x}Ca_{x}Ba_{2}Cu_{3}O_{7 - \delta}$ thin films. By fitting our
data to the theory of Tanaka and Kashiwya\cite{Kashiwaya:1995}, we
are able to find the symmetry of the PP at different barrier
transparencies. Our results are best fitted to the model using a
PP having the \emph{d}\textit{+is} symmetry. The amplitude of the
$is$ component is large at low Z ($Z<0.5$), where it reaches 80\%
of the \emph{d} amplitude, and small at large Z. This dependence
on the barrier transparency implies that the large value of the
$is$ component seen at low Z values is the result of an unusual
proximity effect, which modifies the PP in the HTS near the
interface with a normal metal. Our results are consistent with the
small imaginary component observed by STM (as a low transparency
contact) as reported by Sharoni et al.\cite{Sharoni:2002}
\begin{figure}[t]
\resizebox{!}{3.5 cm}{\includegraphics{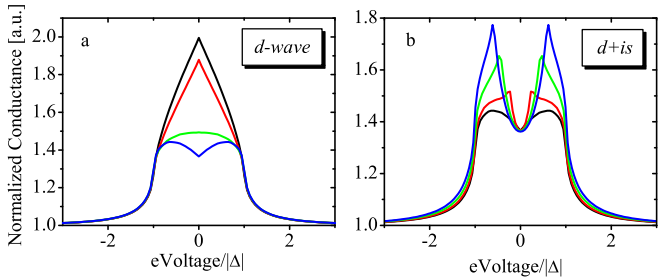}}
\caption{\label{FIG.1}(color in online version) Theoretical
calculation(T=0):normalized conductance vs normalized voltage.(a)
\emph{d} PP, (100) contact,
Z=0(black),0.1(red),0.4(green),0.5(blue)(b)$\emph{d}+is$ PP, (100)
contact,
z=0.5,$\frac{\Delta_{s}}{\Delta_{d}}$=0(black),0.25(red),0.5(green),0.75(blue)
$|\Delta|$ is the maximum amplitude of the PP.}
\end{figure}
\begin{table}[t]
\caption{\label{tab:table1}Summary of fit parameters and X ray
data .}
\begin{ruledtabular}
\begin{tabular}{ccccccccccc}
 No.&Ca&Xray\footnotemark[1]&$Tc_{o}$\footnotemark[2][K]&
 $Tc_{d}$\footnotemark[3][K]&$Z$&$\Delta_{d}$\footnotemark[4]&$\Delta_{s}\footnotemark[4]$&$\Gamma$\footnotemark[4]\\
\hline 1&0& (100) & 91 & 89 &0.34
& 20.1 & 16.3 & 0.9 \\
2&0& (110) & 86 & 78 &0.39
&  13.0 & 10.7 & 0.9 \\
3&0.05& (110) & 88 & 77 &0.40
& 16.3 & 12.0 & 1.7 \\
4&0.2&(110)&76&64&0.46
&15.9 &8.5&1.7\\
5&0.1& (001) & 91 & 82 &0.49
& 17.1 & 7.2 & 1.8 \\
6&0.05&(001)&90&85&0.52
&25.1&0&1.4\\
7&0& (110) & 83 & 63 &0.53
& 11.4 & 7.2 & 3.2 \\
8&0.1& (110) & 91 & 86 &0.60
& 19.4 & 0 & 2.1 \\
9&0.1& (100) & 79 & 71 &0.68
& 16.6 & 0 & 2.8 \\
10&0.1& (001) & 85 & 71 &0.68 &18.6 &0.8 &2.7 \\
11&0.05& (110) & 88 & 77 &0.70
& 18.1 & 1.3 & 6.1 \\
\end{tabular}
\end{ruledtabular}
\footnotetext[1]{(100) orientation was used in the fit for all
contacts}
 \footnotetext[2]{Transition onset}
\footnotetext[3]{Transition downset}
 \footnotetext[4]{meV}
\end{table}
\begin{figure*}[t]
\resizebox{!}{5 cm}{\includegraphics{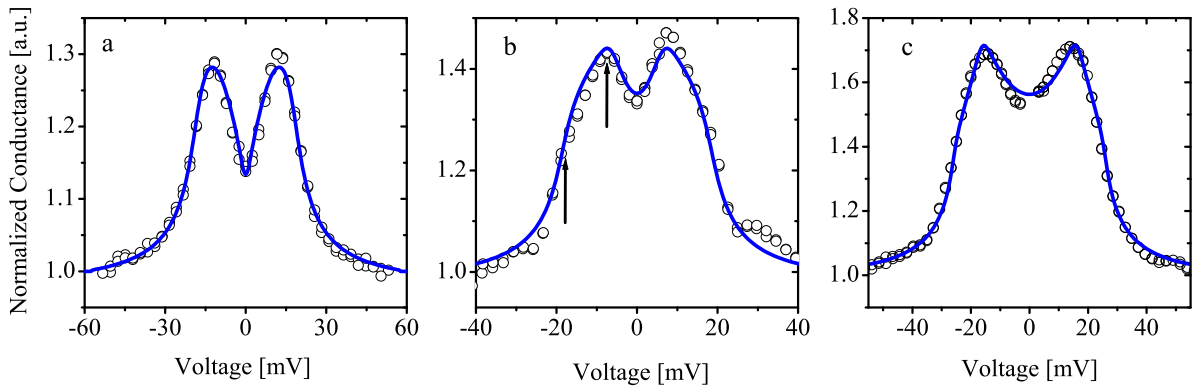}}
\caption{\label{FIG.2}Normalized conductance vs. Voltage; T=4.2 K
(circles) Theoretical fit (line).(a)Sample 10,$R_{N}=40
\Omega$,$\Delta=18.6meV\cos(2\theta)+i0.8meV,Z=0.68,\Gamma=2.7
meV$ and (100) orientation(b)Sample 5,$R_{N}=24
\Omega$,$\Delta=17.1 meVcos(2\theta)+i7.2meV, Z=0.49,
\Gamma=1.8meV$ and (100) orientation. Arrows point out to the
manifestation of the two energy scales, maxima and change of
slope(c)Sample 1,$R_{N}=4.8 \Omega$,$\Delta=20.1
meV\cos(2\theta)+i16.3 meV, Z=0.34, \Gamma=0.9 meV$ and (100)
orientation \texttt{}.}
\end{figure*}

In this study we have used $Y_{1-x}Ca_{x}Ba_{2}Cu_{3}O_{7-\delta}$
sputtered thin films, with x=0,0.05,0.1 and 0.2. The growth
procedure of these films were described in ref.
\cite{Dagan:2000,Sharoni:2003}. The critical temperature of the
films was determined by measuring the resistance versus
temperature. X-ray diffraction was used to determine the global
film orientation (table 1). Sharoni \emph{et
al.}\cite{Sharoni:2002}, using STM topographic scans have shown
that (110) films grown under the same conditions as those we have
used, expose (100) facets. SEM and AFM pictures of the Ca doped
(001) films show a-axis grains on the surface of the film. Thus,
in all three film orientations (100), (110) and (001), (100)
facets are exposed at the film surface. Ozawa \emph{et
al.}\cite{Ozawa:1997} have studied the electronic properties of
the surfaces of (110) oriented YBCO films. They have found that
the degradation time of (110) facets is significantly smaller than
that of (100) facets. We therefore expect that in our films the
chances of obtaining a good metallic contact with a (100) facet
are considerably higher in comparison to a (110) facet. Our
contacts were formed using a mechanically cut Au tip mounted on a
differential screw. All measurements were taken at a temperature
of 4.2K.

We have analyzed the measured conductance spectra by fitting them
to theoretical curves calculated using the formulas developed by
Tanaka\cite{Kashiwaya:1995}. The procedure requires selection of
the PP symmetry. We have tried the following possibilities:
\emph{d}, $\emph{d}+is$, $\emph{d}+id_{xy}$ and $\emph{d}+s$.
Within the selected symmetry the fitting algorithm has the
following adjustable parameters : the amplitudes of the PP
components, the barrier strength Z, the orientation of the surface
in contact with the normal metal ((100), (110) etc.) and the
contact's degree of directionality, namely the width of the
tunnelling cone. The latter is dependent upon the barrier
strength, Z. Following Wei \emph{et al.}\cite{Wei:1998}, in a high
Z, tunnelling contact, we would expect the width of the cone to be
around $20^{o}$ while for a clean N/S contact we expect a value of
almost $90^{o}$. Therefore in calculating the theoretical fitting
curves, for our clean N/S contacts, we have used a $90^{o}$ cone.
The temperature for all calculated curves was set to 4.2K in
accordance with the experimentally measured value. A life-time
broadening parameter (Dynes\cite{Dynes:1978}) $\Gamma$ was used to
account for any smearing beyond the thermal one. We have found all
our results to be best fitted using a $\emph{d}+is$ symmetry and
(100) orientation, adjusting only the values of $\Delta_{d},
\Delta_{s}, Z$ and $\Gamma$(see table 1). Some of the curves can
be fitted also using a $\emph{d}+id_{xy}$ symmetry PP, however
this requires using a narrow tunnelling cone in the fit, which is
unreasonable for a metallic, low barrier, contact. As for the
$\emph{d}+s$ PP symmetry, it requires using
$\Delta_{s}>\Delta_{d}$, thus, we find it less probable, as it
fails to explain experimental data obtained for high Z
contacts\cite{Sharoni:2002,Dagan:2001} and would suggest that the
s-wave channel is stronger than the d-wave one, which is in
conflict with the findings of most experimental data
\cite{Tsuei:2001}.

In Fig. 2 we show three examples of our measured conductance data
and the best fit curve for each of them. They are ordered by
increasing contact's transparency, as determined by the Z value
obtained from the best fit. The absence of a ZBCP suggests that
for a \emph{d} PP the data could be fitted only assuming a (100)
oriented contact. Fig. 2a shows a maxima in the normalized
conductance with a value of around 1.3 which one would expect
could be fitted using the \emph{d} symmetry.(see Fig. 1a). Indeed,
the best fit parameters are given by Z=0.68,
$\Delta=18.6meV\cos(2\theta)+i0.8meV$,$\Gamma=2.7 meV$ and (100)
orientation. In this case the additional $is$ component needed to
fit the data is very small and considering the smearing factor
$\Gamma=2.7 meV$ could very well be even zero. Fig. 2b shows two
distinct features (marked by arrows in the figure), first a maxima
at around $\pm$7.4 mV, and second a distinct change of slope at
around $\pm$18 mV. It is impossible to reproduce this behavior for
the case of a pure \emph{d} PP, as it has only a single energy
scale, but we can correlate these experimental features to the
values of $\Delta _{s}$ and
$|\Delta|\equiv\sqrt{(\Delta_{d})^{2}+(\Delta_{s})^{2}}$
respectively in the case of a $\emph{d}+is$ PP (see Fig. 1b). For
this contact, we find $Z=0.49$, and the amplitude of the $is$
component is enhanced. The other fit parameters are given by
$\Delta=17.1 meV\cos(2\theta)+i7. 2 meV$,$\Gamma=1.8meV$, and
(100) orientation, giving $\frac{\Delta_{s}}{\Delta_{d}}\approx
0.4$. Fig. 2c shows the conductance curve measured on our highest
transparency contact, $Z=0.34$. The amplitude of the maxima in the
normalized conductance is around 1.7. Comparing to Fig. 1a, it
obviously can not be fitted using the pure \emph{d} PP (a peak
amplitude of 1.7 can be reached using $Z\approx3$ but then the
zero bias conductance would be lower than the high bias one).
Using the $\emph{d}+is$ PP we can reproduce the higher peak
amplitude without lowering the zero bias value (see Fig. 1b). We
obtain a best fit using $\Delta=20.1 meV\cos(2\theta)+i16.3 meV$,
$\Gamma=0.9 meV$, and (100) orientation. As can be judged from
Fig. 1b the main effects of adding an \emph{is} component are to
increase the split between the conductance peaks and to enhance
the maximum conductance. This trend is apparent in Fig. 2a trough
Fig. 2c.
\begin{figure}[t]
\resizebox{!}{5 cm}{\includegraphics{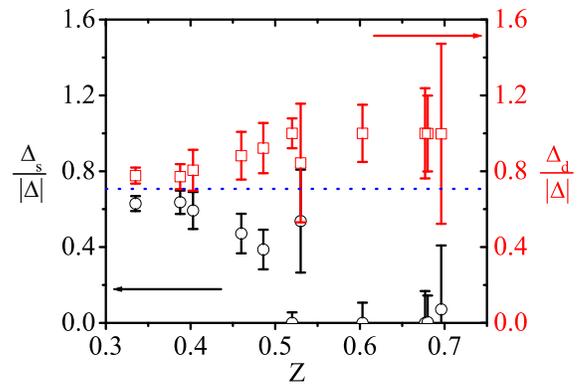}}
\caption{\label{FIG.3}The relative weight of the PP components,
$\frac{\Delta_{s}}{|\Delta|}$(circles) and
$\frac{\Delta_{d}}{|\Delta|}$ (squares) as a function of the
barrier strength, Z. The low Z value to which the two normalized
PP components appear to converge, $\frac{1}{\sqrt{2}}$ (dotted
line)  }
\end{figure}

\begin{figure}[b]
\resizebox{!}{4 cm}{\includegraphics{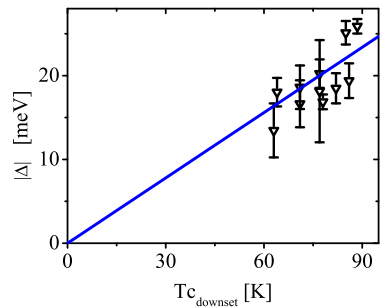}}
\caption{\label{FIG.4} Maximum amplitude of PP,
$|\Delta|\equiv\sqrt{(\Delta_{d})^{2}+(\Delta_{s})^{2}}$ as a
function of $Tc_{downset}$ (triangles)}
\end{figure}
In Fig. 3 we have plotted $\frac{\Delta_{s}}{|\Delta|}$ and
$\frac{\Delta_{d}}{|\Delta|}$ as a function of Z. As the samples
have different critical temperatures we can expect variations in
the values of $\Delta _{d}$ and $\Delta _{s }$ even for a constant
value of Z. To eliminate these changes we have normalized the two
PP components, $\Delta _{d}$ and $\Delta _{s}$, by dividing them
by $|\Delta|$. The relative amplitude of the \textit{is} component
increases from values as low as 0-10{\%} for Z around 0.7 up to
values of around 60{\%} for Z around 0.3. At the same time the
relative amplitude of the d-wave component decreases. There is no
obvious correlation with the level of Ca doping or with $T_{c}$.
The slope of $\frac{\Delta_{s}}{|\Delta|}$ as a function of Z is
maximum at $Z \approx 0.5$. At low Z, the two normalized PP
components appear to converge towards a value close to
$\frac{1}{\sqrt{2}}$, i.e. $\Delta_{s} \approx \Delta_{d}$. We
thus see a change from a PP which is an almost pure d-wave for the
high Z contacts to a PP with almost equal amplitudes of the two
components in the low Z regime, the crossover between the two
regimes occurring around $Z=0.5$. In Fig. 4 we have plotted
$|\Delta|$ as a function of $Tc_{downset}$. The data is consistent
with $2|\Delta|=\eta K_{B}T_{c}$, where $\eta=6.0 \pm 0.4$. The
fact that there is some scattering in the data is to be expected
as the films are never completely homogeneous. One must remember
that $|\Delta|$ represents a local property of the film (at the
contact), whereas T$_{c}$ measures a global property.

We have shown that our data corresponds to a \emph{d}\textit{+is
}symmetry PP, with the value of the imaginary component being a
decreasing function of Z. A development of a subdominant imaginary
PP (SIPP) near the surface of a d-wave SC was predicted
theoretically by Tanuma \emph{et al.}\cite{Tanuma:2001} who
preformed a self consistent calculation of the spatial dependence
of the PP. They did not predict any change in the PP symmetry for
a (100) contact. For the case of a (110) contact in which the
predicted SIPP is maximal, the lower is the contact's
transparency, the stronger is the reduction of the d-wave
amplitude and the higher is the SIPP amplitude, i.e. the SIPP
amplitude is predicted to be a decreasing function of the
contact's transparency. This is inconsistent with our experimental
findings. However, Tanuma \emph{et al.} did not take into account
the possibility that an induced pairing amplitude can appear in
the N side, i.e. a proximity effect, and the related effect this
may have on the PP in the SC in the vicinity of the interface. The
proximity effect between a HTS and a normal metal was studied by
Y. Ohashi\cite{Ohashi:1996} who predicted that a s-wave symmetry
pairing amplitude is induced in N by the \emph{d} SC. According to
his prediction this should lead to a reduction of the \emph{d}
amplitude towards the N/S interface on the superconducting side.
This effect is maximized for a contact with the (100) face of a
\emph{d} SC and is enhanced in high transparency contacts.

We believe that our findings can be explained as a result of a
proximity effect by making the following change in the proposal of
Y.Ohashi. Indeed when a normal metal is in contact with a (100)
boundary of a d-wave SC there appears a s-wave symmetry pairing
amplitude in the normal metal and a decrease of the d-wave
amplitude in the HTS near the boundary. But in order to reduce the
loss of condensation energy on the S side, we suggest that a
\textit{is }SIPP develops in the d-wave SC, in the vicinity of the
barrier, in a way similar to that predicted by Tanuma \emph{et
al.}\cite{Tanuma:2001} for a (110) oriented contact. Considering
our experimental findings, we conclude that YCaBaCuO has a
subdominant s-wave pairing channel, otherwise the appearance of
the \textit{is} PP would have been energetically quite
unfavorable. This subdominant channel exists for a broad range of
Ca doping. Comparing our results to those of Gonnelli \emph{et
al.}\cite{Gonnelli:1}, obtained on LaSrCuO, we find that though
the symmetry used by Gonnelli \emph{et al.} is a different one
(\emph{d}\textit{+id}$_{xy}$ compared to the $\emph{d}+is$) both
experiments indicate that an imaginary component is added to the
\emph{d} PP. All of the spectra reported by Gonnelli \emph{et al.}
are for high transparency contacts ($Z<0.5$) therefore no Z
dependence of the two components can be inferred from their
results. However the ratio of
$\frac{\Delta_{d_{xy}}}{\sqrt{(\Delta_{d_{xy}})^{2}+(\Delta_{d_{x^{2}-y^{2}}})^2}}\approx0.6$
that they find for doping values around the optimum level is not
too far from the one we report here for the same doping levels in
YCaBaCuO. It is important to point out that though we conclude
that the large \textit{is }component that we have measured for
highly transparent contacts is a result of an unusual proximity
effect, we do not suggest that this is the only scenario leading
to the appearance of a SIPP in HTS. Various groups have measured a
split in the ZBCP, explained by an appearance of a SIPP, in both
planar\cite{Krupke:1997,Covington:1997} and STM
\cite{Sharoni:2002} tunnelling junctions with (110) orientation,
where the low transparency suggests that the mechanism leading to
the appearance of the SIPP is unrelated to a proximity effect. A
SIPP was also found by Farber and Deutscher\cite{Farber:1} from
measurements of the temperature dependence of the penetration
depth in Ca over-doped YBCO samples, where no normal metal is
involved.

\begin{acknowledgments}
This work was supported in part by the Heinrich Hertz-Minerva
Center for High Temperature Superconductivity, the Israel Science
Foundation and by the Oren Family chair of Experimental Solid
State Physics. The authors thank Roy Beck for implementing the
computer algorithm used to fit the data and fruitful discussions.
\end{acknowledgments}

\end{document}